\documentclass[11pt,a4paper]{JHEP3}

\usepackage{epsfig}
\usepackage{cite}
\usepackage{listings}

\usepackage{amssymb,amsfonts}

\def\CH{{\cal H}}
\def\CL{{\cal L}}
\def\CN{{\cal N}}\def\CP{{\cal P}}

\def\CW{{\cal W}}

\def\p{\partial}

\def\Tr{{\rm Tr}}

\def\a{\alpha}
\def\d{\delta}\def\e{\epsilon}

\def\l{\lambda}
\def\m{\mu}\def\n{\nu}
\def\r{\rho}\def\s{\sigma}
\def\t{\tau}

\newcommand{\vev}[1]{\left\langle{#1}\right\rangle}

\newcommand{\be}{\begin{eqnarray}}
\newcommand{\ee}{\end{eqnarray}}
\newcommand{\nn}{\nonumber}

\title{ Instanton Partons in   5-dim  $SU(N)$  Gauge Theory  }
\author{
Stefano Bolognesi$^a$ and Kimyeong Lee$^b$ \\
$^a$Department of Applied Mathematics and Theoretical Physics, \\
University of Cambridge, UK \\ 
$^b$Korea Institute for Advanced Study,\\ Seoul 130-722, Korea \\
{\tt s.bolognesi@damtp.cam.ac.uk, klee@kias.re.kr}
}

\abstract{ The circle   compactification of   the 6-dim (2,0) superconformal theory of $A_{N-1}$ type   leads  the 5-dim $SU(N)$ maximally supersymmetric  gauge theory. Instanton solitons embody Kaluza-Klein modes and are conjectured to be composed of partonic constituents.  We realize such a parton of   $1/N$ instanton topological charge   at the intersection of magnetic flux sheets. After a further compactification with  nontrivial Wilson line expectation value, instantons or calorons have been shown to be split into fundamental monopoles of fractional instanton charge.   In the symmetric phase with trivial Wilson line expectation value,  BPS instanton partons emerge more concretely as nonabelian BPS monopoles of minimum charge allowed in Dirac quantization.}

\begin{document}
 

\section{Introduction  and Concluding Remarks }

The   6-dim (2,0) superconformal theory has arisen as the decoupling limit on multiple M5 branes   as well as the low energy limit of   type IIB string theory on  ADE type singularities~\cite{Witten:1995zh}.   Its nonabelian nature and $N^3$ degrees of freedom on $N$ M5 branes have remained mysterious so far~\cite{Klebanov:1996un,Henningson:1998ey}. Its low energy dynamics upon the compactification on a circle leads to the 5-dim $\CN=2$  Yang-Mills theory. Surprisingly, instantons play the role of Kaluza-Klein (KK) modes\cite{Seiberg:1996bd,Rozali:1997cb}. There has been some recent work arguing that the this 5-dim   theory, although apparently nonrenormalizable,  may contain all the degrees of freedom of  the 6-dim (2,0) theory and so finite~\cite{Douglas:2010iu,Lambert:2010iw}.

An instanton  in $SU(N)$ gauge theory has $4N$ zero modes instead of the  4 zero modes which are expected for the position of a single particle. This led to a speculation that a single instanton in $SU(N)$ theory is made of $N$ instanton partons of  $1/N$ instanton topological charge~\cite{Fateev:1979qf,Berg:1979cq,Belavin:1979fb}.  On the other hand, the fractional KK momentum states have appeared before on multiply wound string on a circle and have played an important role in the study of black holes~\cite{Das:1996ug,Maldacena:1996ds}. Similarly multiple D4 branes on top of each other could be regarded as a single M5 wrapping a circle several times.  Such interpretation might allow  the existence of  $1/N$ fractional KK momentum states in 5-dim theory,  which can be regarded as  instanton partons. The black hole solutions for multiple D4 branes were used to show that the second order transition occurs at the temperature at instanton parton mass, indicating the instanton partons are real object~\cite{Itzhaki:1998dd}. However, the exact nature of such instanton partons has remained quite elusive so far, even after a considerable effort to identify them~\cite{Collie:2009iz,Tong:2010zz}.   The standard lore is that there is no localized classical field configuration in $R^4$ which has the $1/N$ instanton charge.

In this work, we show that such instanton partons  can appear more concretely in several places. First we show that   they appear  naturally at the intersection of magnetic flux sheets in 5-dim gauge theory   when the  magnetic flux is the minimum amount undetectable by adjoint matters.     After   compactification on a circle with nontrivial Wilson line,   instantons  or  calorons are known to be  decomposed to fundamental monopoles of fractional instanton charge~\cite{Lee:1997vp}. This 5-dim theory compactified on a circle is  the (2,0) theory compactified on two torus.   The complex coupling of 4-dim is the moduli parameter of the torus and the S-duality of the 4-dim theory is the moduli transformation of the torus. The Wilson loop would be a manifestation  of the non-trivial `Wilson  surface' of nonabelian 2-form field on torus. When this Wilson loop is trivial, we construct here   BPS instanton partons as nonabelian BPS monopoles by providing their asymptotic behavior outside their ``quantum core''.

For any localized physical system, one would expect that the total KK momentum or instanton number to be integer as we require the periodicity under the circle direction. Instantons are at best dipole-like and have localized. However if one consider a system extended to infinity, the reason might be compelling. Our construction instanton partons requires the field configuration to be extended to spatial infinity either by magnetic flux sheet or magnetic charge.  
 
Instantons or calorons in the $SU(N)$ maximally supersymmetric gauge theory $R^{1+3}\times S^1$ with nontrivial Wilson line are composed of the fundamental constituents, $N$ kinds of  magnetic monopoles, which carry  fractional instanton number~\cite{Lee:1997vp,Lee:1998bb,Kraan:1998kp}. The mass of a given fundamental monopole would take a fractional value of the instanton mass which  depends on the Wilson line expectation value.  Classically each fundamental magnetic monopole  has four zero modes.     A single caloron or instanton would be made of $N$ distinct fundamental monopoles. All these monopoles and calorons preserve half-supersymmetry and so are BPS objects. Especially when the  Wilson line  becomes trivial, $N-1$ distinct monopole become  massless and form an nonabelian cloud around the remaining   massive monopole~\cite{Harrington:1978ve,Lee:1996vz}.

Merons~\cite{deAlfaro:1976qz,Callan:1978bm} are another example of fractional instanton number. 
Merons come  only with half instanton number and have log divergent energy.
They cannot be   instanton partons we are looking for as they have infinite energy in large distance. They have divergent attractive forces among them, and so they are not BPS. Moreover there is no obvious way to construct a $1/N$ version of the meron solution for $N \geq 3$.

  We do not know how to describe or quantize the 6-dim (2,0)   theory in the symmetric phase. It is a purely quantum theory as three form tensor field is selfdual. After the circle compactification, we have the 5-dim gauge theory with dimensionful coupling constant.   Of course we should include instantons as nonperturbative objects, which should include instanton partons. It is somewhat easier to approach (2,0) theory in the Coulomb phase, in which case however instanton partons disappear   as   M5 branes get separated from each other. After all the gauge bosons in the 5d theory in Coulomb phase is made of massless abelian   
  
Instanton partons would appear only in the symmetric phase.  In $R^{1+4}$, they should always come together for any localized physical states such that the total topological charge is an integer. We have classical instanton configurations which are imagined to be made of partons. If a single instanton is regarded to have   $N$ parton constituents, the $4N$ instanton moduli space would originate from the spatial and gauge group  coordinates of its spatial position and also its orientation. Thus each of instanton parton would have only four zero modes which should specify its spatial coordinate and gauge algebra coordinate.  Clearly it is a bit hard
task. As only instantons appear in the localized physical states, one may just need to define their group orientations relative to each other's.  Note that the zero modes corresponding to the group orientation of a single instanton is due to the `global part' of nonabelian gauge group, not the local orientation which depends on the local gauge transformation. 
 
Here we show that instanton partons appear  classically in 5-dim at the   intersecting point of two magnetic flux sheets. The gauge orientation of the instanton parton is fixed by the gauge orientation of the flux sheets.
While we assume that $N$ distinct instanton partons form a single instanton, it is not clear how to characterize it even in the long distance. We do not have definite idea about the topological charge of the gauge field on three sphere with zero field strength except for the
$SU(2)$ case, in which case the half winding is allowed.

As we do not have definite progress in flat 5-dim, we compactify the theory on one more circle, ending up with the (2,0) theory compactified on two torus with trivial Wilson line. Fortunately instanton partons seem to appear more clearly after this.   Each instanton partons upon compactification on a circle has a  long-distant profile which is BPS and has $1/N$ instanton topological charge. They have nontrivial magnetic profile whose Dirac string cannot be observed by any   matter in the adjoint representation. This is a more elaborated version of the fact that the regular 't Hooft-Polyakov monopole configuration has the twice large charge than what is required by Dirac quantization   with adjoint matter only.

Our point of view on instanton parton in $R^{1+3}\times S^1$  is that it is somewhat similar to magnetic monopoles in 4-dim maximally supersymmetric gauge theory. Monopoles  arise as solitons in weak coupling limit. Their classical field   profile, which is abelian and BPS, is correct outside 
the core-region, which is the Compton wave length of W-boson. Its BPS central charge determines its mass, which is identical to the  classical energy obtained by integrating the energy density from the spatial infinity to the core region. Inside the core region, the field profile is nonabelian and also quantum in the sense that it is less than the W-boson Compton length. For instanton partons in $R^{1+3}\times S^1$ in the symmetric phase, we  construct the BPS classical field profile outside the  core region of the Compton length size of another KK mode partons along the new circle. Such instanton partons can  exist because of the nontrivial magnetic flux profile at the spatial infinite. Even  though it is defined by a diagonal part of the unbroken gauge group, there is no instability as it is locally BPS.

Let us consider the instanton partons  as the fractional KK modes on the $N$ M5 branes wrapping torus with trivial holonomy or trivial Wilson line expectation value. There are several ways to wrap the torus. Depending on which, one could have the $1/N$ KK momentum mode along either direction of torus. Naturally the fractional KK modes would get excited at the energy   scale which is the minimum of the one-$N$-th fraction of the KK momentum along both compactification directions. The $x^5$ direction is the monopole direction and $x^4$ is the electric direction. Similar to the fact that a monopole core size is that of $W$-boson Compton length,   the core size of the instanton parton is $N$-times larger than the radius of the electric direction circle.

From the perspective that the $SU(N)$ (2,0) theory is comon two torus the physics is gets complicated. S-duality which is the moduli transformation on the four-dimensional theory plays an important role.

The core of a instanton parton in $R^{1+3}\times S^1$ is   $N$ times larger that the naive Kaluza-Klein scale. This is consistent with the expectation that the fully quantum mechanical 6-dim theory including all instanton partons are effective at the energy scale above one-$N$-th fraction of the KK momenta along both compactified directions.

The idea that instanton partons are playing an important role in the strongly coupled regime and are crucial in color confinement has been developed extensively~\cite{Zhitnitsky:2006sr,Toublan:2005tn,Parnachev:2008fy}. It seems that instantons can not provide an explanation for confinement. Both strong coupling and topology must be taken into consideration for the confinement problem in 4dim. 
Our instanton partons as Euclidean time object  may be also possible if there is no fundamental matter field, or all colors fields are arranged so that only adjoint matter fields are appearing. It is not clear how our explicit construction of instanton partons would help in making some progress along this direction.

Our way to construct instanton partons by using the center  $Z_N$ of the group $SU(N)$ could not be generalizable to other simple laced groups of $DE$ types as their center group is much smaller. It would be interesting to find a way to overcome this deficiency.

One of the key issues on M5 branes is what are the fundamental degrees of freedom   as the counting of the degrees of freedom in gravity or anomaly calculation leads to $N^3$.  For the ADE series, the number is argued to be exactly the product of the size of the group and the dual Coxeter number~\cite{Harvey:1998bx,Intriligator:2000eq,Yi:2001bz}. The number of zero modes for a single instanton is four times the dual Coxeter number, implying the number of instanton partons per a single instanton is that number again. Collie and Tong~\cite{Collie:2009iz,Tong:2010zz} have proposed an intriguing idea that the degrees of freedom could be accounted  if the instanton partons somehow belong to the adjoint representation. Our point of view is somewhat different. We think that instead 1/4 BPS
junctions of selfdual strings in the Coulomb phase should play a role in the counting, and recently we have provided some supporting argument which shows the number of 1/4 BPS states in the Coulomb phase matches one-third of the product of the dual Coxeter number and the dimension of the group for all ADE cases~\cite{Bolognesi:2011rq}.

The reality of instanton partons can be settled only when we understand the symmetric phase of the (2,0) theory wrapped on a circle. Our approach is a step forward to give a definite picture of instanton partons.  Some sort of exact calculation involving the (2,0) theory on circles or torus is needed to settle the issue.

The plan of this work is as follows. In Sec.2, we review 5-dim gauge theory and show some fractional winding is possible. Also instanton partons are shown to appear at  the intersection point of magnetic flux sheets. In Sec.3, we review the compactification of the (2,0) theory on two torus. In Sec.4, we explore the KK modes around two torus when there is nontrivial Wilson-line expectation value. The constituents are  magnetic monopoles and charged W-bosons. Sec. 5,  we describe the BPS field profile for  the instanton partons in 5-dim gauge theory on $R^{1+3}\times S^1$ at the symmetric phase.

\section{ \boldmath$ {A_{N-1} }$ type  (2,0) theory on \boldmath${R^{1+4}\times S^1}$}

The ADE (2,0) theory has the self-dual 3-form field strength and so no adjustable coupling constant~\cite{Witten:1995zh}.  There is no natural nonabelian generalization of 2-form gauge field and no classical Lagrangian. Thus the (2,0) is supposed to be defined only quantum mechanically.  The $A_{N-1}$ type theory arise from $N$ M5 branes and has a simple low energy description upon a circle compactification 
\be x^5\sim x^5+2\pi R_6.  \ee
The low energy dynamics of the $A_{N-1}$ (2,0) theory upon circle compactification becomes     the 5-dim $\CN=2$ $SU(N)$ supersymmetric gauge theory with  the Lagrangian 
\be \CL_5= \frac{1}{2g_5^2} \Tr\, F_{MN}F^{MN} + \cdots , \ee
where $M, N=0,1\cdots 4$  and dots are for fermionic and scalar contributions. The instantons are KK modes of the compactification, and so its mass is  identified with the KK momentum, leading to the relation
\be M_{instanton}= \frac{8\pi^2}{g_5^2}=\frac{1}{R_6} . \ee
In abelian (2,0) theory in 6-dim, there are self-dual three form field strength.  The 5-dim gauge field and field strength can be identified as $A_M \sim B_{M5}$ and $F_{MN}\sim  H_{MN5}$. The momentum density along $x^5$ is proportional to $-H_{0\m\n }H_{5\m\n} \sim -F_{\m\n}\tilde F_{\m\n}$, which is the topological charge  density.

A general instanton configuration satisfies selfdual equation $F_{\mu\nu}= \tilde F_{\mu\nu}\equiv \e_{\mu\nu\r\s}F_{\r\s}/2$ where $\mu,\n=1,2,3,4$, and its mass becomes
\be \CP_5 =-\frac{1}{2g_5^2}\int d^4x \Tr F_{\mu\nu}F_{\mu\nu} =-\frac{8\pi^2}{g_5^2} \n ,
\label{6mt}\ee
 where instanton charge is the Pontryagin index
\be \n = \frac{1}{16\pi^2 } \int d^4 x \, \Tr F_{\mu\nu} \tilde F^{\mu\nu} = \frac{1}{8\pi^2 }\int d^4 x \, \p_\m W_\mu ,   \ee
where  Chern-Simons current is
\be W_\mu =   \e_{\mu\nu\rho\s} \Tr (A_\nu \p_\r A_\s -\frac{2i}{3} A_\n A_\r A_s) .
\ee
The gauge field becomes pure gauge at the boundary  $A_\mu=i U^\dagger \partial_\mu U$ and so the instanton charge becomes the winding number 

\be \n(U) = \frac{i}{24\pi^2}\int_{S^3_\infty} \Tr (   U^\dagger dU)^3 , \label{windn}\ee
which characterizes the asymptotic behavior of the gauge field.
When the function $U$ is a well-defined map from three sphere to nonabelian gauge group, the winding number of the gauge at the spatial infinity is the third homotopy group which is $\pi_3(G) =Z$. One can choose the gauge where the gauge field becomes trivial at spatial infinity but 
has some singularities in finite region which is characterized by nontrivial winding numbers.

A single instanton in a semi-simple gauge group $G$ has $4h_G$ zero modes where $h_G$ is the dual Coxeter number. We assume that a single instanton  consists of  $h_G$ partons of equal topological charge and four zero modes~\cite{Bernard:1977nr}. Instanton parton is expected to be BPS and has the topological charge,
\be \CP_6 = \frac{8\pi^2}{g_5^2}\frac{1}{h_G}.  \ee
Table I shows the dual Coxeter number $h_G$ and the center group for  semi-simple Lie  group $G$.
\vskip 1cm 
\begin{center}
\begin{tabular}{|c|c|c|}
\hline  
$G$ & $h_G$ & center \\
\hline  
$A_{N-1}$ & $N$ & $Z_N$ \\
$B_N$ & $2N-1$  &  $Z_2$\\
$C_N$ & $N+1$  &  $Z_2$  \\
$D_{2N}$ & $4N-2$   & $Z_2\oplus Z_2$ \\
$D_{2N+1}$ & $4N$  & $Z_4$ \\ 
$E_6$ & 12  & $Z_3$ \\
$E_7 $ &  18 & $Z_2$ \\
$E_8$ & 30  & \{1\} \\
$F_4$ & 9  & \{1\} \\
$G_2$ &  4  &  \{1\} \\
\hline  
\end{tabular}
\vskip 1cm 
 \centerline{\footnotesize Table I: the dual Coxeter number and the center of the universal covering group $G$}
\end{center}

We will see  later that instanton partons can be realized easily for the $A_{N-1}=SU(N)$ gauge theories by using its rather large center group. On the other hand, our method does not easily generalized to other simple-laced groups.  To get a single instanton parton in $SU(N)$ case, we regard  that the gauge field configuration is sort of classical in large spatial region and is characterized by the fractional winding number.   In the theory with matters in   adjoint representation only, the
gauge function $U: S^3_{\infty}\rightarrow SU(N)$ can be multi-valued modulo the center of the gauge group $Z_N$,  which leaves $A_\mu=i U^\dagger \p_\mu U$ single-valued. One hopes to find such $U(x)$ such that    the winding number $\n$ fractional, say, $1/N$.  Unfortunately, we could not find such map except for $N=2$ case at the moment and it is not clear whether such multi-valued map $U(x)$ exists for  $N\ge 3$.

To see half-winding map for $N=2$ case, consider   the Euler angles $\theta,\varphi,\psi$ on $S^3_{\infty}$ by using the map from the three sphere to $SU(2)$: 
$g=e^{i\varphi \t_3/2}e^{i\theta \t_2/2}e^{i\psi \t_3/2} $. 
The relation 
$-i \bar g dg = \frac{\tau_i}{2} \s_i $ defines the left-invariant 1-forms 
\be && \s_1 = \cos\psi\sin\theta d\varphi-\sin\psi d\theta , \nn  \\
&& \s_2=\sin\psi\sin\theta d\varphi+\cos\psi d\theta , \nn \\
&& \s_3=d\psi+ \cos d\varphi  .
\ee
The metric on the $S^3_\infty$ is
\be ds^2=\frac14\Big( d\theta^2 + \sin^2\theta d\varphi^2+ (d\psi+\cos\theta d\varphi)^2 \Big) .\ee
The volume form is
\be dV=\frac18 \s_1\wedge \s_2\wedge \s_3 = -\frac18 \sin\theta d\theta \wedge d\varphi\wedge d\psi . \ee
Now the range of $\theta,\varphi,\psi$ is
\be 0\le \theta\le  \pi,\   0\le \psi+ \varphi  \le 4\pi, \ 0 \le \psi-\varphi\le 4\pi , \ee
which is equivalent to 
\be 0\le \theta\le  \pi,\   0\le   \varphi  \le 2\pi, \ 0 \le \psi \le 4\pi .
\ee

Let us choose a double-valued map $U_2$  from $S^3_{\infty}$  to $SU(2)$ which is given as
\be U_2 &=& e^{i\varphi\t_3/2}e^{i\theta\t_2/2}e^{i\psi\t_3/4} \nn \\
&=&   \left(\begin{array}{cc} +\cos\frac{\theta}{2} e^{ +i\varphi /2} &
+\sin\frac{\theta}{2} e^{+i \varphi /2} \\
-\sin\frac{\theta}{2}e^{-i \varphi /2} & +\cos\frac{\theta}{2} e^{-i \varphi /2}   \end{array}\right)
 \times {\rm diag}( e^{+i\psi/4}, e^{-i\psi/4}). \ee
This  is locally well-defined but double-valued under $\psi\rightarrow \psi+4\pi$ by $ -1$. The gauge field $A_\mu \simeq i\bar U\p_\mu U$ is single-valued and its winding number $\n=1/2$. If we consider only adjoint matter field, such a gauge transformation is allowed in principle, implying the possibility of  BPS half-instantons.
The very same solution $A_\mu = i\bar U\p_\mu U$ considered everywhere is a good one, and is very similar to ordinary instanton with zero size. This is pure gauge everywhere a part from the singularity at the origin that contains all the $F \widetilde{F}$ charge.    
The generalization to $SU(N)$ for a $1/N$ winding would probably be impossible. The previous construction relies on the existence of the Hopf fibration for $S^3/S^1 = S^1$ and for $SU(N)$ not all the circles passing trough a point are equivalent, and only a discrete set are the ones passing trough the center of the group. 
It is of course possible to embed the previous map in $SU(N)$ since a minimal winding is always an $SU(2)$ embedding, but this would still give a $1/2$ fractional map.

Other object in the literature with fractional instanton number is the merons were again only the $1/2$ fraction is realized.
This object was first introduced in \cite{deAlfaro:1976qz}. The BPST instanton solution is $A_{\mu} = {\eta_{\mu\nu}} x_{\nu}/(|x|^2 + \rho^2)$ where $ \eta_{\mu\nu}$ are 't Hooft tensor matrices and $\rho$ is the instanton scale. This self dual solution breaks the conformal symmetry leaving $8$ parameters zero modes. The meron solution is very similar in $A_{\mu} =  \eta_{\mu\nu} x_{\nu}/(2|x|^2)$ where the size parameter is zero and the $1/2$ factor is the crucial difference. Although the decay of the gauge potential is $A \sim 1/x$ as for the instanton, the field strength leading term is $F \sim 1/x^2$ and not $F \sim \rho^2 / x^4$. The meron is thus log divergent, both in UV and IR. It carries $1/2$ instanton charge and is clearly not self-dual. Note that the instanton charge is completely hidden in the singularity, the field strength outside is nonzero but has vanishing charge.  The UV singularity can be smoothed out by a $\rho$ scale, although this is no longer classical stationary point. The $F \sim 1/x^2$ behavior at infinity, like monopoles in three dimensions, was  suggested could be used to explain confinement in four dimension \cite{Callan:1978bm} with a generalized Polyakov mechanism. The log divergence in the action  would be overcome for sufficiently strong coupling like the BKT phase transition in two dimension.
The meron can also be embedded in gauge theories with higher $N$. The instanton charge still remains $1/2$ (the embedding does not affect this) and thus it cannot be considered the instanton partons for $N>2$. 
 There is some evidence that this mechanism could indeed be active in $SU(2)$ \cite{Lenz:2003jp} but, due to the fact that merons are not in general related with the center of the gauge group for $ N\geq 3$, is unlikely that this is the mechanism for confinement for general $N$.
The fact that the previous two constructions give only fractional instanton charge $1/2$, suggests that there should be a relation between them but at the moment is not clear. 


We started requesting our instanton parton to be BPS, smooth and with $1/N$ fractional instanton charge.
Even forgetting about the first two conditions, we have not even showed so far the possibility of $1/N$ fraction of charge. We show it now that isolated $1/N$ fractional instanton configuration in $R^4$  appear naturally when one considers  two intersecting magnetic flux sheets (or surface operators).  Let us start by considering    two magnetic flux sheets on $1-2$ and $3-4$ planes with trivial holonomy in the fundamental representation. Their non-vanishing 
magnetic fluxes are, for example,
\be &&  F_{12}= 2\pi {\rm diag} (1, 0 , \cdots, 0) \delta(x^1)\delta(x^2) , \nn \\
&& F_{34} = 2\pi  {\rm diag}( 1, 0,\cdots , 0)\d(x^3)\d(x^4)  . \ee
Their vector potential is
\be A_\mu dx^\mu =  {\rm diag} \Big( d {\rm Arg}(x^1+ix^2)+ d {\rm Arg}(x^3+i x^4) , 0, 0 , \cdots, 0\Big) . \ee
The magnetic flux is such that such volume operators cannot be detected by the matter in the fundamental representation.   
Such configuration is not BPS. The two sheets meet at the origin. The topological charge at the 
meeting point is
\be \n = 1, \ee
which is a single instanton.
Instead if there are  only adjoint matter fields, the undetectable magnetic flux would have the trivial holonomy modulo the group center. We choose two nontrivial magnetic flux sheets intersecting on the origin with magnetic flux to be
\be &&  F_{12}= \frac{2\pi }{N}{\rm diag} (N-1,-1,\cdots,-1,-1) \delta(x^1)\delta(x^2),  \nn \\
&& F_{34} =\frac{2\pi }{N} {\rm diag}( 1, -N+1, 1,\cdots , 1, 1)\d(x^3)\d(x^4) ,  \ee
whose vector potential is
\be A_\mu dx^\mu &=& + \frac{1}{N} {\rm diag} (N-1,-1,\cdots,-1,-1)d{\rm Arg}(x^1+ix^2) \nn \\
& & +  \frac{1}{N} {\rm diag}( 1, -N+1, 1,\cdots , 1, 1) d{\rm Arg}(x^3+ix^4) .
\ee
Such configuration is not BPS. The two sheets meet    at the origin.  (Such flux sheets are  sometimes named as center vortices and have been argued to play the key role in confinement~\cite{tHooft:1978cv}.)  

The topological charge at the 
meeting point is easily evaluated to be
\be \n  = \frac{1}{N} ,
\ee
which is that of an  instanton parton. When one considers the lattice gauge theory with only adjoint matter field, such configurations could be relevant. 
In the continuum instead this configuration is singular. We could smooth it out, making that magnetic flux supported on an area greater than zero. But this would lead to  a  classically unstable configuration. In short, it is difficult to  describe BPS instanton parton configurations in $R^4$, even if we restrict the BPS condition and ask only smoothness and $1/N$ fractional charge. It seems a no-go, although we do not have enough information to state a precise theorem. The only way seems to concentrate all the fractional instanton charge in a point-like singularity.

\section{ \boldmath$ {A_{N-1}}$ type (2,0) theory on \boldmath$ {R^{1+3}\times T^2}$}

Once we compactify the (2,0) on a circle, let us proceed to compactify it further on a torus so that
\be (x^4 , x^5) \sim (x^4, x^5+2\pi R_6), \ \ (x^4, x^5) \sim (x^4+2\pi R_5, x^5-\theta R_6 ). \ee
While we do not know the (2,0) theory, we naively expect that the  Kaluza-Klein modes on the torus are characterized by the momentum factor
\be \exp \Big\{  i n_4 \frac{  x^4}{R_5} + i n_5 \Big( \frac{x^5}{R_6}+\frac{\theta x^4}{2\pi R_5} \Big) \Big\}  . 
 \ee 
The four dimensional mass for a given mode, ignoring the interaction, would be 
\be  M(n_4,n_5) = \frac{1}{R_5} | n_4 + \tau n_5|  = \sqrt{ \frac{1}{R_5^2}\Big(n_4+\frac{\theta n_5}{2\pi}\Big)^2 + 
\frac{n_5^2}{R_6^2} } . \label{mn45}\ee
(Here we are imagining  a massless abelian scalar field in (2,0) theory whose field equation is $(-\p_0^2 + \sum_{i=1}^5\p_i^2)\Phi=0$.)

Let us consider the effect of the $SL(2,Z)$ modular transformation of the torus. 
 Under the $SL(2,Z)$ transformation
\be \tau \rightarrow \tilde \t = \frac{a\tau +b}{c\tau+d} ,\ee
where $a,b,c,d \in Z$ and $ad-bc=1$, 
we transform the torus to
\be (x^4, x^5)\sim (x^4, x^5+ 2\pi R_6 |c\t +d| ), \ (x^4, x^5)\sim (x^4+  \frac{2 \pi R_5}{|c\t +d|}, x^5- 2\pi R_6 |c\t +d|{\rm Re}\tilde \t       ).
\ee
The momentum factor of a single-valued KK mode is transformed to   
\be \exp\Big\{ i\tilde n_4 \frac{|c\t +d| x^4}{R_5} + i\tilde n_5\big( \frac{x^5}{R_6|c\t+d|} + \frac{|c\t+d| {\rm Re}\tilde \t x^4}{R_5} \big)\Big\} ,
\ee
and its free field mass is transformed to  
\be \tilde M_{\tilde n_4,\tilde n_5} = \frac{|c\t +d|}{R_5} | \tilde n_4 +\tilde \t \tilde n_5| . \ee
With the transformation of the quantum numbers
\be  (\tilde n_5, \tilde n_4) = (n_5, n_4) \left(\begin{array}{cc} d & -b \\ - c & a\end{array}\right) ,
\ee
we get the invariance of the mass term, 
\be \tilde M( {\tilde n_4,\tilde n_5} ) =M(n_4,n_5). \ee

From the 5-dim gauge theory point of view, the above $SL(2,Z)$ transformation becomes a bit different and also specific. The low energy 4-dim Lagrangian is
\be \CL_4&=& -\frac{1}{ 2g_4^2} \Tr F_{mn}F^{mn}  -\frac{1}{g_4^2}\Tr F_{m4}^2  -\frac{\theta}{16 \pi^2} \Tr F_{mn} \tilde F^{mn} +\cdots \nn \\
&=& -  \frac{1}{16\pi}   {\rm Im} \, \Tr \Big\{ \tau (F_{mn} + i \tilde F_{mn})^2\Big\}  -\frac{1}{g_4^2} \Tr F_{m4}^2 + \cdots , \ee
where $m,n =0,1,2,3$ and $\tilde F^{mn}=\frac12 \e^{mnpq}F_{pq}$ with $\e^{0123}=1$ and the dots denote the contributions from scalars, fermions and also all     the Kaluza-Klein modes of $x^4$ compactification. 
The four-dim coupling constant is
\be \frac{4\pi}{g_4^2} = \frac{8\pi^2 R_5}{g_5^2}= \frac{R_5}{R_6}, \label{4coupling}\ee
and the 4-dim complex coupling parameter
\be \tau = \frac{\theta}{2\pi} +\frac{4\pi i}{g_4^2} \ee
is the complex structure of the two torus.
The conjugate momentum density is
\be \Pi_i = \frac{2}{g_4^2}F_{0i}-\frac{\theta}{4\pi^2} B_i, \ee
where
\be B_i =\frac12 \e_{ijk} F_{jk} ,\ee
and the Gauss law is
\be D_i \Pi_i+\cdots =0 ,  \ee
where the dots denote the contributions from matters and KK modes.
 
The 5-th directional KK modes are instantons. 
The linear momentum (\ref{6mt}) along $x^5$ is
\be \CP^5=  -\frac{2}{g_5^2} \int d^4x \Tr B_i F_{i4} =\frac{n_5}{R_6},\ee
where the Pontryagin index $\n=-n_5$ and $8\pi^2/g_5^2=1/R_6$. 
The linear momentum   along $x^4$ is proportional to
\be \CP^4= - \frac{2}{g_5^2}\int d^4x \Tr  F_{0i}F_{4i} = -\frac{1}{2\pi R_5} \int d^4x\Tr \Pi_i F_{4i} + \frac{\theta}{2\pi}\frac{R_6 \CP_6}{R_5}  .  \ee
The quantization should be
\be    \CP^4=\frac{1}{R_5}\Big(n_4+\frac{\theta n_5}{2\pi} \Big) , \ee
where
\be n_4 = -\frac{1}{2\pi}\int d^4 \Tr \Pi_i F_{4i} .\ee
Here we are ignoring all the contributions from other matter fields.
The  energy functional can be written as
\be \CH 
&=& \frac{1}{g_5^2}\int d^4x \Tr \Big( F_{0i}^2 + F_{04}^2 + B_i^2 + F_{4i}^2\Big) \nn \\
&=& \frac{1}{g_5^2}\int d^4x \Tr \Big( (F_{0i} -\sin\a F_{i4})^2   +  ( B_i-\cos\a F_{i4})^2 + F_{04}^2\Big)  \nn \\
& & \ \ + \CP^4\sin\a + \CP^5\cos\a  , \ee  
and so the BPS bound is given by the mass (\ref{mn45}): 
\be \CH \ge \sqrt{(\CP^4)^2+ (\CP^5)^2} = M(n_4,n_5)  .\ee
The BPS equation becomes $F_{04}=0$ and
\be F_{0i}=F_{i4}\sin\a, \ B_i= F_{i4}\cos\a .\ee
While we do not explore the detail of the supersymmetry, all KK modes are 1/2 BPS and the quantum states should such that they are all transforming to each other under the $SL(2,Z)$ transformation. 
Note that $SL(2,Z)$ duality, which is certainly there for the 6-dim theory compactified on a two torus, is not necessarily expected a priori for the 5-dim theory compactified on a circle.  It would of course be a natural fact if the conjecture equivalence between the 5-dim and 6-dim theories would be true~\cite{Douglas:2010iu,Lambert:2010iw}.

As we have compactified the 5-dim theory on a circle, the gauge invariant
  Wilson-loop  
\be \CW = \Tr \CP \exp i\int dx^4 A_4   \ee
can have nontrivial expectation value. There are allowed large gauge transformations which is multivalued but leaves the adjoint matters single-valued. An example is
\be U(x^4)=\exp\Big( \frac{ix^4}{R_5N } {\rm diag}(N-1,-1,-1,\cdots,-1)\Big).\label{largeg} \ee
The above transformation is single-valued modulo the center group $Z_N$.  
The expectation value of $A_4$ can be arranged by such large gauge transformation and Weyl reflections so that
\be \langle A_4\rangle = \frac{1}{R_5}(h_1,h_2,\cdots h_N), \ee
 such that $\sum_a h_a=0$, 
 \be   h_1\ge h_2\ge \cdots \ge h_N \ge h_1-1 .     \ee
When the gauge symmetry is broken completely  to abelian subgroups, all expectation values are different, $h_1>h_2> \cdots > h_N > h_1-1  $.  After T-duality on $x^4$ circle, the values $h_a/R_5$ can be interpreted as the position of D3 branes in the dual circle of radius $1/(2\pi R_5)$.

The Wilson line expectation value is the degrees of freedom for the virtual nonabelian flux through the $x^4$ circle as it is a generalization of the  
   Aharonov-Bohm phase.   From the (2,0) theory of point of view, we have now torus and two-form tensor gauge field. While we do not know the detail nonabelian formalism, the above Wilson-line expectation value captures the physics of the nonabelian Aharonov-Bohm effect due to the virtual three-form flux.  The gauge symmetry is spontaneously broken by the `Wilson-surface'.  

\section{W-bosons and Monopoles as Fractional KK Momentum Modes }

As we consider the case of the (2,0) theory compactified on a two-torus with nontrivial Wilson-line,
the gauge symmetry is spontaneously broken. First of all, there would be massive charged vector bosons
whose electric charge is quantized. By the Gauss law, the canonical momentum is quantized as
\be \Pi_i =\frac{2}{g_4^2}F_{0i} \sim  \frac{ 2x_i Q_e}{4\pi r^3}  , \ee
where
\be Q_e= \frac12 {\rm diag}(q_1, -q_1 + q_2,-q_2,\cdots, -q_{N-2}+q_{N-1}, -q_{N-1}) .\ee
where $q_i$ are integers.  Each charged vector boson for the roots of the broken generators are
quantum mechanically BPS, denoting fundamental strings connecting two D3 branes in the T-dual picture. While the configuration can arise only quantum mechanically, its abelian profile is BPS so $F_{0i}=F_{i4}$ and 
\be A_4 \sim \vev{A_4}- \frac{g_4^2}{4\pi  r}Q_e.\ee
The energy for W-boson connecting $a$ and $a+1$ brane would be obtained by the quantization.
Another way is just to calculate its BPS charge from the spatial infinity and convert it to the energy   
\be (\CP^4)_a = \frac{2}{g_4^2}\int d^3x \p_i \Tr F_{0i} A_4 = \frac{h_a-h_{a+1}}{R_5} .  \ee
  Here the W-boson mass is the fractional $\CP^4$ momentum.     There would be another $x^4$-dependent $W$-boson which can be identified as the fundamental string connecting $N$-th and $1$-st D3 branes and would have mass and charge
\be && (\CP^4)_N= \frac{h_N-h_1+1}{R_5} , \\
&& Q_e= \frac12 (-q_N, 0 \cdots, 0 , q_N). \ee
The total momentum along $x^4$ would be
\be \CP^4 = \sum_{a=1}^N q_a (\CP^4)_a  , \ee
and the total electric charge would be
\be Q_e =\frac12 {\rm diag}(-q_N+q_1,-q_1,\cdots -q_{N-1}+q_N). \ee
A single KK mode along $x^4$ direction would be the sum of these $N$ fundamental W-bosons which carries zero electric charge and linear momentum $1/R_5$. This is the  fundamental string wrapping once around the dual $x^4$ circle in the type II B picture. There would be fundamental strings connecting any pair of D3 branes after wrapping the dual circles many times which correspond to the KK modes with higher $x^4$ momentum. The Wilson loop symmetry breaking shows how such modes with fractional momentum along $x^4$ direction appear.

 Let us now review the magnetic monopoles and calorons which appear the modes for fractional  $x^5$ KK momentum~\cite{Lee:1997vp,Lee:1998bb,Kraan:1998kp}.
   The time independent fundamental monopole solutions give by the solution of the BPS equation
\be B_i =\frac12 \e_{ijk}F_{jk} = D_i A_4 -\p_4 A_i \sim \frac {  x_i Q_m}{r^3} .\ee
The asymptotic form of the magnetic field and the scalar field can be transformed to be abelian:
\be B_i = \frac{   x_i Q_m }{r^3} , \ \  A_4= \langle A_4 \rangle -\frac{Q_m }{r} .\ee
The well-known smooth BPS monopole configuration describes the nonabelian core region.  
The magnetic charge is given by  
\be Q_m = \frac12{\rm diag} (p_1, -p_1+p_2,-p_2+p_3,\cdots, -p_{N-2}+p_{N-1}, -p_{N-1} ).\ee
Each of the $a$-th fundamental magnetic monopole carries also electric charge due to the Witten effect
an so the electric field becomes
\be \frac{2}{g_4^2}F_{0i} \sim \frac{2  x_i Q_e}{4\pi r^3} \sim \frac{\theta}{4\pi^2} B_i  , \ee
where the electric charge is
\be Q_e = \frac{\theta}{2\pi}Q_m .
\ee
Thus the a-th fundamental magnetic monopole has the instanton number
\be (\CP^5)_a =-\frac{8\pi^2}{g_5^2}  \n=  -\frac{2}{ g_5^2}  \int d^4 x \p_i \Tr (B_i A_4)    = -\frac{h_a-h_{a+1}}{R_6} \label{mm}.\ee
Thus they carry fractional $x^5$ KK momentum. In addition, the fractional electric charge means they carry  electric charge proportional to $\theta$, or the fractional $x^4$ KK momentum proportional to $\theta$, leading to the mass
\be m_a = \frac{h_a-h_{a+1}}{R_5} |\tau | .\ee
Each fundamental magnetic monopole has four zero modes for its position and phase, and is interpreted as $D$ string connecting $a$ and $a+1$-th  D3 branes in type IIB picture.

In addition there is a time-dependent fundamental monopole solution which arises from the KK mode of charge
\be  Q_{KK}  = ( -p_N, 0  \cdots, 0 ,0,p_N). \ee
The BPS configuration is obtained by first making the large gauge transformation and get a new
 the new expectation value
 \be \vev{A_4}_{new}=\frac{1}{ R_5} {\rm diag}( h_1-\frac{N-1}{N}, h_2+\frac{1}{N},\cdots  h_N+\frac{1}{N}).
 \ee
The BPS configuration in the abelian region is the similar magnetic field and the 4-th gauge field becomes    
 \be A_4= \vev{A_4}_{new}  -\frac{Q_{KK}}{r} .\ee
Its instanton charge or $x^5$ linear momentum is 
\be (\CP^5)_N= -\frac{h_N -h_1+1}{R_6}.  \ee
In this gauge the KK fundamental monopole is time-independent and has again four zero modes.
With a large gauge transformation, one goes back to the original Wilson line value, and the KK monopole has the $x^4$ time dependence in the core region.  
A single instanton is made of $N-1$ distinct fundamental BPS monopoles and a single fundamental KK BPS monopole and the $x^5$ KK linear momentum t is
\be (\CP^5)_{instanton} = \sum_{a=1}^{N } p_a(\CP_5)_a   = -\frac{1}{R_6} . \ee
The total magnetic charge vanishes and so calorons or instantons on $R^3\times S^1$ are magnetic dipoles at best. There would no force between these $N$ fundamental monopoles and they could regarded as   constituent parts of calorons.

Let us consider the decompactification limit $R_5\rightarrow \infty$, The nonabelian core size of each monopole would be of order $R_5$. If their separation scale is $D$ is fixed, the core of these monopoles would overlap each other, and it would be hard to separate from each other. 
 The analysis of a single caloron  in $SU(2)$ case shows that the distance $D$ between monopoles and the instanton scales $\rho$ are related in the decompactification limit as
\be \rho^2 \sim  R_5 D .\ee
In order to get a finite scale instanton configuration in infinite $R_5$, we need to put the distance between monopole component to vanish. When there are multi-instantons, the generalization of the above scaling law is not clear as there are many scales in the problem. 

When one analyze the superpotential for the 4-dim pure $\CN=1$ gauge theory to find out the gluino condensation, one see that the natural object to give such contribution is not instanton, but instanton partons. We believe that this aspect is not still well-understood in 4-dim.  When one compactifies this theory on a circle, the magnetic monopole constituents of instantons lead to the extended Toda-type superpotential. Due to the presence of the  adjoint fermion field, the potential has the minimum when all constituent monopoles have the same action, the $1/N$ of instanton action~\cite{Davies:1999uw}.  However, this is a dynamical process 
of minimization, and is highly dependent on the matter content. See~\cite{Diakonov:2007nv,Unsal:2007jx} for the more recent development along this direction.

\section{Instanton Partons }

Instead of having a nontrivial Wilson line with a circle compactification $x^4\sim x^4+2\pi R_5$, 
let us consider the symmetric phase with the trivial Wilson line,
\be \vev{A_4} ={\rm diag}(0,0,.\cdots 0).\label{symm1}  \ee
A single caloron discussed in the previous section now can be regarded as constructed of   one massive KK monopole and $N-1$ massless monopoles. Massless monopoles are not gauge invariant and would form a nonabelian cloud around the KK monopole. Depending on the size of the cloud, the configuration interpolates the BPS monopole solution and the singular instanton~\cite{Harrington:1978ve,Lee:1996vz}.  In addition,  KK modes of the gauge field carrying the momentum $\CP_4$ would arise quantum mechanically and they would describe massive particles in the adjoint representation of the unbroken gauge group $SU(N)$.  

We argue here that  there exist a new class of objects in this symmetric phase.  In the symmetric phase of $N$ M5 branes wrapping a torus, another interpretation is possible:  a single M5 brane wrapping the torus $N$ times.
 There is several different ways to realize this which are characterized by two natural numbers $(w_4,w_5)$ such that $w_4 w_5=N $. Here a single M5 brane wraps $x^4$ cycle $w_4$ times and $x^5$ cycle $w_5$ times. The fractional momentum along $x^4$ and $x^5$ may be possible    in principle with
\be &&  \CP^4 = \frac{1}{R_5}( \frac{n_4}{w_4} + \frac{\theta n_m}{2\pi w_5}) , \\
&& \CP^5= \frac{n_m}{w_5 R_6} . \ee
%
Here we just consider the case where $(w_4=1,w_5=N)$. In this case the object would carry just $1/N$ instanton charge.

We want to imagine that a single instanton is made of $N$ instanton partons which present in $R^4$ space and the group space. Each of instanton partons are supposed to be BPS and have 4 zero modes. The $4N$ zero modes of a single instanton may denote the position variables of $N$ instanton partons in spatial and group coordinates. Instanton partons are also regarded as a   KK modes for $1/N$ KK momentum on a single M5 wrapping the $x^5$ circle $N$ times. Thus we think that instanton partons are BPS objects of $1/N$ instanton mass and four zero modes living only in the symmetric phase of the gauge group. In the phase with nontrivial Wilson line, all instanton partons would be confined.  After a further circle compactification on $x^4$ circle and nontrivial Wilson line, instantons or calorons would be decomposed to fractional instantons or monopoles.    
 
There are several  gauge equivalent ways to express this symmetric phase (\ref{symm1})  and 
we   use a large gauge transformation (\ref{largeg}) 
to transform the above vacuum to
\be \vev{A_4} =\frac{1}{N R_5}{\rm diag}(N-1,-1, \cdots -1) .\label{symm2}\ee
They denote the position of D3 branes on the dual circle of circumference $1/R_5$ and 
one can see that both of the above two vacua denote the configuration where $N$ 
D3 branes are on top of each other. 

We  imagine that a fundamental BPS instanton parton appears as a BPS magnetic monopole 
in large distance, even though the gauge symmetry is not broken to
abelian subgroup.   The second vacuum (\ref{symm2}) appears as if the gauge symmetry is broken partially.  We are imagining a magnetic monopole configuration which is allowed only with matters in adjoint representation. The magnetic charge would be smaller than the case where larger  that the case where there is a fundamental representation, as our electric probe in the adjoint representation is cruder.  
Monopole configuration in large distance is given by the Wu-Yang construction,  
\be 
A_i dx^i= \left\{ \begin{array}{cc}
     A_i^U dx^i = +Q(1-\cos\theta)d\varphi   ,  &  \theta < \frac{\pi}{2}+\epsilon   \\
 A_i^D dx^i = -Q(1+ \cos\theta)d\varphi  , &   \theta > \frac{\pi}{2} -\epsilon 
 \end{array}\right. \ \ .
 \ee 
The gauge transformation between upper and lower hemisphere is
\be A^D = G  A^U \bar G -i dG \bar G ,  \ee
where  
\be G= e^{-2i Q\varphi} .\ee
We require the above configuration to be BPS, which means that $F_{ij} = \e_{ijk}F_{k4}$. As $F_{ij} = \e_{ijk} Q x_k/ r^3$, we get the 4-th component of the gauge field to be
\be A_4= \vev{A_4}-\frac{Q}{r}. \ee

 When there exists the matter in the fundamental representation, we
require that $G$ to be single valued.   A simple case which has single-valued transition function 
is the case of a single caloron with magnetic charge  
\be Q_{instanton} =\frac{1}{2}{\rm diag}( 1, -1, 0 , \cdots , 0 ) .  \ee
One can calculate the topological charge to be that of instanton, 
\be \CP^5= \frac{16\pi^2 R_5}{g_5^2} \Tr( Q A_4) =\frac{8\pi^2}{g_5^2}=\frac{1}{R_6} .\ee
The amount of the charge is that of the smooth BPS monopole configuration with nonabelian core. This monopole is gauge equivalent to   a limit of a single caloron configuration in the symmetric phase where the nonabelian cloud has been sent to infinity.

As we assume all the D4 branes are on the top of each other, all  finite energy open strings have both of their ends on D4 branes. Thus there is only adjoint matter field and so the mathematically allowed transition function $G$ could be in the center of the $SU(N)$ group. For example we can choose the charge to be
\be Q_{parton} = \frac{1}{2N}{\rm diag}(1, -N+1, 1, \cdots 1). \ee
The transition function $G$ is multi-valued by the group center. Such charge is acceptable as the 
Dirac quantization is satisfied by only adjoint matter fields. The BPS configuration for $A_4$ is
\be 
 A_4= \vev{A_4} -\frac{Q}{r} =\frac{1}{N R_5}{\rm diag}\Big( N-1-\frac{R_5}{2r}, -1+\frac{R_5}{2r}(N-1), -1-\frac{R_5}{2r},\cdots ,-1-\frac{R_5}{2r} \Big) . \nn \\
\label{partonconf} \ee
As one moves into the core from the spatial infinity,  $A_4$ breaks the gauge symmetry to $U(1)\times SU(N-1)$ and then again   reaches again the symmetric region  when $r\approx R_5 /2$: 
\be A_{4}(r=2R_5) \approx \frac{1}{NR_5} {\rm diag} (N-2, N-2, -2,\cdots , -2) .
\ee
This partially broken region seems to be strange to have. Indeed the quantum core should be reached earlier. 
  The topological charge from the spatial infinity would be 
\be \CP^5   = \frac{16\pi^2 R_5}{g_5^2} \Tr \ Q \vev{A_4}  = \lim_{r\rightarrow\infty} \frac{8\pi^2}{Ng_5^2} \Big(1-\frac{R_5}{2r}(N-1)\Big) =\frac{8\pi^2}{Ng_5^2} = \frac{1}{NR_6}. 
\ee

For the magnetic monopoles, the field energy outside the nonabelian core saturates the energy bound, implying the inner  region is nonabelian, and quantum in the sense it is of the Compton wave length of $W$ bosons. For the instanton partons, let us again assume that the BPS energy is exhausted from the field energy outside the core region $r>R_c$.  As
\be \frac{16\pi^2 R_5}{g_5^2} Tr (Q A_4)(r=R_c) = \frac{8\pi^2}{N g_5^2} ( 1-\frac{R_5}{2R_c}(N-1))=0 , \ee
we see the quantum core region is 
\be r \lesssim R_c = \frac12 (N-1) R_5. \ee
This value is at the boundary roughly  the gauge symmetry $SU(N)$ is broken  to $U(1)\times SU(N-1)$.   One would say the BPS field configuration is valid outside the core region
\be r \gtrsim R_c\approx N R_5, \ee
and so locally stable. 
Instanton parton configuration given here carries almost all energy outside the core region. Inside the core region, the field configuration (\ref{partonconf})   should not be trusted. We expect  new degrees of freedom or quantum feature shows up at the core scale.

This is consistent with the following view. The original theory is 6-dim and is UV finite. First is compactified on a circle $R_6$ and then on a circle $R_5$. Let us assume that instanton partons and the KK momentum parton along $R_5$ circle become effective at their mass scales $1/(NR_6)$ and $(1/NR_5)$, respectively. Thus, the (2,0) theory compactified on torus is quantum in the length scale satisfying
\be \ell_{quantum} \lesssim {\rm max} (NR_5, NR_6 ). \ee
In other words, the (2,0) theory appears in the energy scale higher than $1/\ell_{quantum}$. 
Our description of the instanton parton in terms of 4-dim theory is valid in the weak coupling limit $g_4^2/4\pi = R_6/R_5 < 1 $  or
\be R_6 < R_5 . \ee
If we have chosen the case $R_5 < R_6$, we have to make the S-dual transformation to get the weak coupling limit, resulting in the previous example.

We need $N$ instanton partons to make a single instanton. We can choose $N-1$ distinct partons whose $a$-th  charge is
\be Q_a = \frac{1}{N}{\rm diag} (1,\cdots, -N+1, \cdots ,1), \ee
where $a+1$-th entry is $-N+1$. They could be regarded as independent of $x^4$ in the gauge where the $A_4$ takes the expectation value (\ref{symm2}).  To make a single instanton, we need $N$ partons of distinct type. The last one is like $KK$-monopole. 
Let us now construct another KK monopole-like parton by considering the vacuum expectation value
\be \vev{A_4}  = \frac{1}{NR} {\rm diag}( -1, N-1,-1,\cdots, -1) , \ee
which is   equivalent to the symmetric vacuum (\ref{symm2}) by a large gauge transformation, 
and consider the magnetic charge
\be Q_{KK} = \frac{1}{2N}{\rm diag} (-N+1,1,\cdots ,1) .
\ee
Again the abelian part of the KK parton configuration is
\be A_4 = \vev{A_4} - \frac{Q_{KK}}{r} = \frac{1}{NR}{\rm diag}( -1+\frac{R}{2r}(N-1),N-1-\frac{R}{2r},-1-\frac{R}{2r},\cdots -1-\frac{R}{2r}) .
\ee
The mass of this KK parton is 
\be \CP^5= \frac{1}{NR_6}. \ee
Of course we have to make a large $x^4$ dependent transformation to make this KK parton to put in the previous symmetric vacuum (\ref{symm2}).  In this vacuum (\ref{symm2}), all  fundamental partons except the KK parton is independent of $x^4$. We think that a single instanton is made of $N-1$ distinct partons of mass $8\pi^2/(Ng^2)$ and magnetic charge $Q_a$ and the KK parton of the same mass and magnetic charge $Q_{KK}$.  Each parton has the core of size $NR_5$. The total magnetic charge vanishes as 
\be \sum_{a=1}^{N-1} Q_a + Q_{KK} = 0. \ee

Now we would make another large gauge transformation to get to the symmetric vacuum $\vev{A_4} =0$. In this obviously symmetric vacuum all partons would be  $x^4$-dependent. While the core-regions of each fundamental partons would $x^4$ dependent, the $B_i$ configuration outside the core regions remain abelian and simple.

The picture emerging here is that a single instanton in $R^3\times S^1$ is made of $N$ partons of equal fractional Pontryagin number  $1/N$. While it is not clear how to count the zero modes for each partons, one could assign four zero modes for its position and phase. One would wonder whether there is additional zero modes for nonabelian transformation. As in the single caloron in the symmetric phase with massless monopoles sent to infinity, there may be some global color problem which does not allow additional internal zero modes, and so we may expect similar global color problem for instanton partons.

To make a single instanton, we need $N$ distinct instanton partons so that the total magnetic charge vanishes. The core size of each instanton partons would be $NR_5$ if $R_6<R_5$. As one puts instanton partons close to each other so that their cores overlap, the Cartan part far outside the core would cancel each other, leaving a dipole moment. Interior region could develop the classical profile of a single caloron in the symmetric phase with finite cloud size.

The instanton partons are different from the monopoles in the previous section as obvious from their charge. The relation can be understood as follows. We have to make a change of variables from the monopole base to the parton base. The monopole base are the simple roots $(1,-1,0,\dots,0)$, $\dots$ , $(0,\dots,0,1,-1)$ plus the one corresponding to the KK monopole $(-1, 0, \dots,0,1)$. The parton base are the $N$ generators $(1, \dots, 1, -(N-1), 1,\dots,1)$. Note that for $N=2$ the parton base is half of the monopole base. This is because the 't Hooft-Polyakov monopoles have twice the charge of the fundamental Dirac monopole which is the one used for the parton.  Clearly the base is redundant because only $N-1$ are linear independent. The fact is that on top of the charge conservation we need also to impose the instanton number conservation. In the symmetric phase the KK monopole carries all the instanton charge, and this fixes that every parton is made with a $1/N$ fraction of the KK monopole.   We thus have the change of basis matrix respectively for $N=2,3,4$
\be
\left( \begin{array}{ll}
0 & 1/2 \\
1 & 1/2
\end{array} \right)\ ,
\quad \qquad
\left( \begin{array}{lll}
-1/3 & 0 & 1/3 \\
\phantom{-}2/3 & 0 & 1/3 \\
\phantom{-}2/3 & 1 & 1/3
\end{array} \right)\ ,
\quad \qquad
\left( \begin{array}{llll}
-1/2 &-1/4 & 0 & 1/4 \\
\phantom{-}1/2 & -1/4 &  0 & 1/4 \\
\phantom{-}1/2 & \phantom{-}3/4 & 0 & 1/4 \\
\phantom{-}1/2 & \phantom{-}3/4 & 1 & 1/4
\end{array} \right) \ .
\ee

The exact nature of these partons in the (2,0) theory compactified on a circle is not clear yet. 
The $4N$ zero modes of a single instanton is made of 4-center of mass and $1$ zero mode for the scale and $4N-5$ modes for the gauge orientation. This   indicates that the position vector  of instanton partons knows the instanton position and its gauge orientation. 
As the ADHM data made of hypermultiplets $(\l_\a, B_\mu)$ where $\l_\a$ is the bi-fundamentals of $SU(N)\times U(k)$ and $B_\m$ are adjoint hypermultiplets of $U(k)$, we could regard the bi-fundamental $\l$ is the data  for the positions of instanton partons.   Further analysis along this direction may yield some insight.

\section*{Acknowledgments}

\noindent We are grateful to David Tong and Pierre van Baal for helpful discussions.
KL is supported in part by NRF-2005-0049409 through  CQUeST,
and the National Research Foundation of Korea Grants NRF-2009-0084601 and NRF-2006-0093850.
SB wants to thank KIAS for the hospitality in March 2011 when part of this work was done.

\end{document}